\documentclass[preprint,preprintnumbers,amsmath,amssymb, superscriptaddress]{revtex4}

\usepackage{textcomp}
\usepackage{makeidx}
\usepackage{amsmath}
\usepackage{subfigure}
\usepackage{amssymb}
\usepackage{hyperref}
\usepackage{graphicx}
\usepackage[utf8]{inputenc}
\usepackage{float}
\usepackage{color}
\usepackage{dsfont}

\begin{document}

\title{Pure electromagnetic-gravitational interaction in Ho\v{r}ava-Lifshitz theory at the kinetic conformal point}

\affiliation{Departamento de F\'isica, Facultad de ciencias b\'asicas, Universidad de Antofagasta, Casilla 170, Antofagasta, Chile.}
\author{Alvaro Restuccia}
\email{alvaro.restuccia@uantof.cl}
\affiliation{Departamento de F\'isica, Facultad de ciencias b\'asicas, Universidad de Antofagasta, Casilla 170, Antofagasta, Chile.}
\author{Francisco Tello-Ortiz}
\email{francisco.tello@ua.cl}
\affiliation{Departamento de F\'isica, Facultad de ciencias b\'asicas, Universidad de Antofagasta, Casilla 170, Antofagasta, Chile.}


\begin{abstract}
We introduce  the electromagnetic-gravitational coupling in the Ho\v{r}ava-Lifshitz framework, in $3+1$ dimensions, by considering the Ho\v{r}ava-Lifshitz gravity theory in $4+1$ dimensions at the kinetic conformal point and then performing a Kaluza-Klein reduction to $3+1$ dimensions. The action of the theory is second order in time derivatives and the potential contains only higher order spacelike derivatives up to $z=4$, $z$ being the critical exponent. These terms include also higher order derivative terms of the electromagnetic field. The propagating degrees of freedom of the theory are exactly the same as in the Einstein-Maxwell theory. We obtain the Hamiltonian, the field equations  and show consistency of the constraint system. The kinetic conformal point is protected from quantum corrections by a second class constraint. At low energies the theory depends on two coupling constants, $\beta$ and $\alpha$. We show that the anisotropic field equations for the gauge vector is a deviation of the covariant Maxwell equations by a term depending on $\beta-1$. Consequently, for $\beta=1$, Maxwell equations arise from the anisotropic theory at low energies. We also prove that the anisotropic electromagnetic-gravitational theory at the IR point $\beta=1$, $\alpha=0$, is exactly the Einstein-Maxwell theory in a gravitational gauge used in the ADM formulation of General Relativity.
\end{abstract}


\maketitle

\section{Introduction}
Although General Relativity has been tested with an amazing physical accuracy, it has not given rise, so far, to a consistent theory of quantum gravity. It can be quantize in the framework of perturbative quantum field theory but it is non-renormalizable. It becomes renormalizable by adding quadratic curvature invariants  to the action but at the cost of violating unitarity \cite{r1}. It may then be considered as an effective theory of a more fundamental theory, with  a much better behaviour at high energies. Superstring theory, a unification theory of all fundamental forces, is such a candidate. However it has not been able to predict fundamental physical aspects as for example at which energy scale the breaking of supersymmetry takes place.

Recently, P. Ho\v{r}ava \cite{r2,r6} proposed a new approach to obtain a renormalizable theory of gravity. The proposal considers an anisotropic scaling, also called Lifshitz scaling, on the space-time:
\begin{equation}
x^{i}\rightarrow bx^{i}, \quad t\rightarrow b^{z}t, \quad i=1,2,3.    
\end{equation}
 
The proposal manifestly breaks the relativistic symmetry. The Ho\v{r}ava action is second order in time derivatives and the potential contains higher order spacelike derivatives up to $z=3$ terms. Following Lifshitz the theory should be power counting renormalizable at the UV regime, moreover Ho\v{r}ava's original idea was that the theory should flow from the UV  to General Relativity at the IR point. 
Several aspects of Ho\v{r}ava-Lifshitz theories have been understood \cite{r4,r5,r7,r8,r9,r10,r11,r12,r13,r14,r15,r16,r17,r18,r19,r20,r21,r22,r23,r24,r25}.  

An important one which remains to be studied is the coupling of Ho\v{r}ava-Lifshitz gravity to the other fundamental interactions, in particular to the electromagnetic interaction. Maxwell equations have been tested with utmost precision. Consequently, if the anisotropic version of them provides a model describing electromagnetism  they should be, at low energies, very "near" Maxwell equations.
Following the Ho\v{r}ava-Lifshitz framework one would like to couple electromagnetism and gravity in a space-time consisting on a foliation with spacelike leaves parametrized by the time variable. The action should contains second order time derivatives and a potential with higher order spacelike derivatives of the geometrical objects of the theory, that is, higher order derivative terms constructed from the 3- dimensional Riemann tensor, the lapse and the electromagnetic field. We will show that the potential must contain at least up to $z=4$ derivative terms in order to have a power counting renormalizable theory.

A natural approach to fulfill these requirements is to consider a Ho\v{r}ava-Lifshitz gravity theory in 4+1 dimensions and to perform a Kaluza-Klein reduction to $3+1$ dimensions. Ho\v{r}ava-Lifshitz geometry is then guaranteed. In \cite{r3} we followed this approach for a Ho\v{r}ava-Lifshitz gravity in $4+1$ dimensions at a non-critical value of the kinetic coupling constant. The theory described the propagation of the pure gravity and pure electromagnetic physical degrees of freedom together with the propagation of two scalars fields. One of them intrinsic to the Ho\v{r}ava-Lifshitz theory at the non-critical value of the kinetic coupling and the other one the dilaton scalar field associated to the Kaluza-Klein approach.
Since we are interested in this work to analyze the propagation of fundamental interactions in the anisotropic framework we would like to avoid the scalar fields. In \cite{r3} the comparison with the Maxwell equations was done in the presence of dynamical scalar fields. In that case, to take the dilaton field on its ground state in the field equations restricts the physical degrees of freedom of the theory. In this paper we consider the pure electromagnetic-gravitational interaction without any dynamical scalar field. The comparison with the Maxwell-Einstein equations can then be done without restricting the physical degrees of freedom. It is a full-fledged comparison. Besides, effects of the presence of the dynamical scalar field characteristic of the Ho\v{r}ava-Lifshitz at the noncritical point establish additional restrictions to the coupling parameters of the theory which we can avoid. For example, the emission of dipolar radiation, due to the presence of a dynamical scalar field, in the orbital evolution of binary pulsars induces a much more rapid decay of the orbital period than the predicted by General relativity, which is in good agreement with observations \cite{n1}. On the other side, in the Horava-Lifshitz gravity at the kinetic conformal point no dipolar radiation exists. Only the quadrupole radiation characteristic of the transverse-traceless tensorial modes is predicted, as in General Relativity \cite{r24}.

An important aspect of the propagation of the electromagnetic and gravitational physical modes concerns with its speed of propagation. The theoretical results should match the recent detection of coincident gravitational and electromagnetic waves \cite{n2,n3}. The propagation speeds should match to within a part in $10^{15}$ \cite{n4,n5}.

Theoretical and experimental data have been used to get upper and lower bounds to the coupling parameters of the Horava-Lifshitz gravity theory, for example \cite{n6,n7,n8,n9,n10,n11,n12}. The parameters $\alpha$ and $\beta$ are severely restricted to be near 0 and 1 respectively, however the dimensionless parameter $\lambda$ is unrestricted at the kinetic conformal point, since no dynamical scalar field is present in the formulation and most of the bounds are related to the presence of the scalar mode. Besides the value of $\lambda$ at the kinetic conformal point is protected of quantum corrections by a second class constraint which restricts the trace of the conjugate momentum of the Riemannian metric on the leaves. Moreover the conjugate momentum is also restricted by a first class constraint, which implies that it is transvers in the sense of York \cite{n13}.The two constraints imply that the conjugate momentum is transvers-traceless. Besides, it follows that this component is independent of the parameter $\lambda$ \cite{n13,n14}. Furthermore a perturbation of the parameter $\lambda$ can only appear through the conjugate momentum. Consequently, the value of $\lambda$ at the kinetic conformal point does not receive quantum corrections.

In this work we analyze the electromagnetic-gravity coupling in the anisotropic Ho\v{r}ava-Lifshitz framework in a formulation where only the pure gravity and pure electromagnetic degrees of freedom propagate. The Ho\v{r}ava-Lifshitz approach to gravity starts by considering a foliated space and time manifold, where each leave of the foliation is parame-
trized by the time coordinate. This a different starting point from the General Relativity one. Since the foliated manifold is given a priori, we have to choose the topology of it. We consider a globally hyperbolic manifold. Any such manifold $\mathcal{M}$ has a smooth embedded three-dimensional Cauchy surface, and furthermore any two Cauchy surfaces for $\mathcal{M}$ are diffeomorphic. Moreover $\mathcal{M}$ is diffeomorphic to the product of a Cauchy surface with $\mathds{R}$ \cite{n15}. We distinguish the Cauchy surfaces which are compact from the ones which are asymptotically flat Riemann surfaces, they are topologically different. We consider in this paper the latter ones. The cosmological spacetimes manifolds, i.e. a spatially compact, globally hyperbolic Lorentzian manifolds, are then excluded from our analysis.

We consider the Ho\v{r}ava-Lifshitz gravity in $4+1$ dimensions at its critical value of the kinetic coupling, the  kinetic conformal point, and perform a Kaluza-Klein reduction to $3+1$ dimensions. The Ho\v{r}ava-Lifshitz  gravity in $3+1$ dimensions at the  kinetic conformal point was analyzed in \cite{r4,r5}. The physical degrees of freedom of the $4+1$ theory are the transverse-traceless gravitational modes in $4+1$ dimensions which decomposes into the transverse-traceless gravitational modes in $3+1$ dimensions, the transverse electromagnetic modes in $3+1$ dimensions and the dilaton field. The intrinsic Ho\v{r}ava –Lifshitz scalar is not present at the  kinetic conformal point. The dilaton being a scalar under diffeomorphisms on the space like leaves of the foliation and reparametrization of time can be fixed in the Hamiltonian without changing the transformation law of the geometrical objects of the theory (the same approach starting with General Relativity in $5$ dimensions give rise to the Einstein-Maxwell theory in $4$ dimensions \cite{r26}). The resulting Hamiltonian in $3+1$ dimensions describes the electromagnetic-gravitational coupling in the Ho\v{r}ava-Lifshitz geometry. The potential contains higher order spacelike derivative terms from the gravitational as well as from the electromagnetic degrees of freedom.

The point of view  we consider in this work, as we did in \cite{r3}, is to interpret the $4+1$ theory and Kaluza-Klein reduction only as a geometrical mechanism to obtain the anisotropic theory in $3+1$ dimensions, which it is expected to be renormalizable in the framework of perturbative quantum field theory. The point of view of considering the Ho\v{r}ava-Lifshitz gravity theory as an effective theory of a more fundamental theory at higher dimensions is also a interesting one, but is not the one we follow here.

The coupling of electromagnetism to Ho\v{r}ava-Lifshitz  gravity at the non-critical values of the kinetic coupling constant and  using a different approach has been studied in reference \cite{r27}. The coupling there is only at the second order in derivatives of the electromagnetic degrees of freedom through the relativistic action of electromagnetism. The power counting renormalizability of the theory is then not guaranteed since no higher order spacelike derivatives of the electromagnetic sector are included in the potential. On the contrary, at least in General Relativity the coupling of relativistic matter makes the situation worst \cite{r28}.

In section \ref{sec2} we consider the $4+1$ Ho\v{r}ava-Lifshitz gravity action at its kinetic conformal point. In section \ref{sec3}, we obtain the Hamiltonian and field equations of the anisotropic electromagnetic-gravitational theory in $3+1$ dimensions after Kaluza-Klein dimensional reduction. Next, in section \ref{sec4} we perform a perturbative analysis in order to cast the physical degree of freedom propagated by theory. In section \ref{sec5}, we analyze the anisotropic electromagnetic field equations at low
energies and compare them to the Maxwell equations. In section \ref{sec6}, we prove that the anisotropic electromagnetic-gravitational theory, at low energies and for a particular value of the coupling constants is exactly the Einstein-Maxwell
theory in a particular gauge used in the ADM formulation of General Relativity.
Finally, in section \ref{sec7} we give our conclusions.

\section{The $4+1$ Ho\v{r}ava-Lifshitz Theory: The Kinetic Conformal Point}\label{sec2}

We start considering the $4+1$ dimensional Ho\v{r}ava-Lifshitz theory. Its action is given by
\begin{equation}\label{eq1}
\begin{split}
S\left(g_{\mu\nu},N_{\rho},N\right)=\int dtdx^{4}N\sqrt{g}\bigg[K_{\mu\nu}K^{\mu\nu}-\lambda K^{2}+\beta\mathbf{R}
+\alpha a_{\mu}a^{\mu}
+\mathcal{V}\left(g_{\mu\nu},N\right)\bigg].  
\end{split}
\end{equation}
It is invariant under diffeomorphisms on the spacelike leaves of the foliation and under reparametrization of the time variable. $K_{\mu\nu}$ stands for the extrinsic curvature of the leaves embedded in the foliation, $K=g^{\mu\nu}K_{\mu\nu}$ denotes its trace, $\mathbf{R}$ is the 4-dimensional Riemannian curvature of the leaves and $a_{\mu}\equiv \partial_{\mu}lnN$. The $a_{\mu}a^{\mu}$ term introduced in \cite{r6} is relevant for the stability of the theory. Due to their presence the second class constraints which appear in the theory becomes elliptic partial differential equations. Only boundary conditions are then required to show existence and uniqueness of their solutions and this fact is relevant in the formulation of the initial value problem. Finally, $\mathcal{V}(g_{\mu\nu},N)$ denotes higher order than two spatial derivative terms which transform as a scalar under diffeomorphisms of the spacelike leaves and as a scalar under reparametrization of the time variable. They are independent of $N_{\mu}$ which plays the role of a Lagrange multiplier in the theory and $\lambda$, $\alpha$ and $\beta$ are coupling constants. The requirement  of renormalizability and that the overall coupling constant of the action must be dimensionless implies that all higher order spatial derivative terms up to $z=4$ must be included in the potential. If the symbol of the potential  for each physical degree of freedom of the theory is elliptic of order $8$ ($z=4$), then the power counting renormalizability of the theory is guaranteed \cite{r5}. The infinitesimal transformation of the fields $g_{\mu\nu}$, $N_{\mu}$ and $N$ under spacelike diffeomorphisms with infinitesimal parameter $\xi^{\mu}$ and time reparametrization with infinitesimal parameter $f(t)$ are 
\begin{eqnarray} \label{eq2}
\delta g_{\mu\nu}&=& \partial_{\mu}\xi^{\rho}g_{\nu\rho}+\partial_{\nu}\xi^{\rho}g_{\mu\rho}+\partial_{\rho}\xi^{\rho}g_{\mu\nu}+f\dot{g}_{\mu\nu} \\ \label{eq3}
\delta N_{\mu}&=&\partial_{\mu}\xi^{\rho}N_{\rho}+\xi^{\rho}\partial_{\rho}N_{\mu}+\dot{\xi}^{\rho}g_{\rho\mu}+f\dot{N}_{\mu}+\dot{f}N_{\mu} \\ \label{eq4}
\delta N&=& \xi^{\rho}\partial_{\rho}N+f\dot{N}+\dot{f}N.
\end{eqnarray}
Notice that the transformation law of $N_{\mu}$ contains the term $\dot{\xi}^{\rho}g_{\rho\mu}$ which characterizes the transformation law of a Lagrange multiplier. Therefore, $N_{\mu}$ does not transforms as a four vector under spacelike diffeomorphisms.

Now, we proceed to formulate the theory in its canonical form \cite{r29,r30,r31,r32}. We notice that only second order time derivatives are present in the Lagrangian. So, the canonical conjugate momentum of $g_{\mu\nu}$, denoted by $\pi^{\mu\nu}$, is given by
\begin{equation}\label{eq5}
\pi^{\mu\nu}=\frac{\partial \mathcal{L}}{\partial \dot{g}_{\mu\nu}}=\sqrt{G}\left(K^{\mu\nu}-\lambda g^{\mu\nu}K\right),   
\end{equation}
and its trace by
\begin{equation}\label{eq6}
\pi=g_{\mu\nu}\pi^{\mu\nu}=\sqrt{g}\left(1-4\lambda\right)K.    
\end{equation}
We notice that $\lambda=1/4$ is a critical point. In fact for that value 
\begin{equation}\label{eq7}
\pi=0,    
\end{equation}
becomes a primary constraint. This value of the dimensionless coupling $\lambda$ depends on the dimensions of the foliation. For a $3+1$ foliation the critical value corresponds to $\lambda=1/3$. The constraint (\ref{eq7}) is the generator of a conformal symmetry of the kinetic term of the Lagrangian, however the terms of the potential break this symmetry. Consequently, (\ref{eq7}) will end up being a second class constraint of the theory. Eq. (\ref{eq7}) and the conjugate momentum to $N$, which we denote by $P_{N}$, equal zero are primary constraints. Explicitly it reads 
\begin{equation}\label{eq8}
P_{N}=0.    
\end{equation}
The primary constraint (\ref{eq8}) arises since no time dependence on $N$ is present in the Lagrangian density. So, the Hamiltonian density is then given by
\begin{equation}\label{eq9}
\begin{split}
\mathcal{H}=N\sqrt{g}\bigg[\frac{\pi^{\mu\nu}\pi_{\mu\nu}}{g}-\beta\mathbf{R}-\alpha a_{\mu}a^{\mu}-\mathcal{V}\left(g_{\mu\nu},N\right)\bigg]
+2\pi^{\mu\nu}\nabla_{\mu}N_{\nu}-\mu\pi-\sigma P_{N},   
\end{split}
\end{equation}
where $\mu$ and $\sigma$ are Lagrange multipliers. Variations with respect to $N_{\mu}$ yields the third primary constraint
\begin{equation}\label{eq14}
\nabla_{\mu}\pi^{\mu\nu}=0.    
\end{equation}
In order to determine the field equations of the theory at low energies, we will not consider the contribution of the higher order derivative terms in the potential. They are crucial in the quantum analysis of the theory and will be included in the Kaluza-Klein reduction to $3+1$ dimensions. So, the field equations besides the primary constraints are

\begin{equation}\label{eq10}
\begin{split}
-\dot{\pi}^{\mu\nu}=-\frac{1}{2}Ng^{\mu\nu}\bigg[\frac{\pi^{\lambda\rho}\pi_{\lambda\rho}}{\sqrt{g}}\bigg]+ 2N\bigg[\frac{\pi^{\mu\lambda}\pi^{\nu}_{\lambda}}{\sqrt{g}}\bigg]   
+\beta\sqrt{g}N\bigg[\mathbf{R}^{\mu\nu}
-\frac{1}{2}\mathbf{R}g^{\mu\nu}\bigg] &\\-\beta\sqrt{g}\bigg[\nabla^{(\mu}\nabla^{\nu)}N -g^{\mu\nu}\nabla_{\lambda}\nabla^{\lambda}N\bigg]
-\frac{1}{2}\alpha\sqrt{g}Ng^{\mu\nu}a_{\rho}a^{\rho}+\alpha\sqrt{g}Na^{\mu}a^{\nu}
&\\+2\nabla_{\rho}\bigg[\pi^{\rho(\mu}N^{\nu)}\bigg]-\nabla_{\rho}\bigg[\pi^{\mu\nu}N^{\rho}\bigg]-\mu\pi^{\mu\nu},
\end{split}    
\end{equation}

\begin{equation}\label{eq11}
\dot{g}_{\mu\nu}=\frac{2N\pi_{\mu\nu}}{\sqrt{g}}+\nabla_{\mu}N_{\nu}+\nabla_{\nu}N_{\mu}+\mu g_{\mu\nu}.    
\end{equation}
On the other hand, conservation of the primary constraints (\ref{eq7}) and (\ref{eq8}) give rise to two new constraints
\begin{equation}\label{eq12}
2\frac{\pi^{\mu\nu}\pi_{\mu\nu}}{g}+\beta\mathbf{R}+\left(\alpha-3\beta\right)a_{\mu}a^{\mu}-3\beta\nabla_{\mu}a^{\mu}=0,  
\end{equation}
\begin{equation}\label{eq13}
H_{N}\equiv\frac{\pi^{\mu\nu}\pi_{\mu\nu}}{g}-\beta\mathbf{R}+\alpha a_{\mu}a^{\mu}+2\alpha \nabla_{\mu}a^{\mu}=0,    
\end{equation}
respectively. Conservation of (\ref{eq12}) and (\ref{eq13}) only determine the Lagrange multipliers $\mu$ and $\sigma$. The Dirac approach to determine all the constraints of the theory ends at this stage. These are all the constraints of the theory.

The Poisson bracket of the constraints imply that (\ref{eq7}), (\ref{eq8}), (\ref{eq12}) and (\ref{eq13}) are second class constraints while (\ref{eq14}) is a first class constraint. 

\section{Kaluza-Klein Dimensional Reduction: From $4+1$ to $3+1$ Dimensions}\label{sec3}

In order to study the coupling between the Ho\v{r}ava-Lifshitz gravity and electromagnetism in an anisotropic framework, we perform in this section the Kaluza-Klein reduction from $4+1$ to $3+1$ dimensions following a pure cylindrical projection. We follow exactly the same procedure explained in detail in \cite{r3}. We consider here, as it was done in \cite{r3} the complete action including the potential terms $\mathcal{V}(g_{\mu\nu},N)$ and assume a topology of the type $\mathcal{M}_{D+1}=\mathcal{M}_{D}\times S^{1}$. We will denote the extra dimension by $x^{4}$ and assume all fields are independent of it. We split the 4-dimensional metric $g_{\mu\nu}$ as  
\begin{equation}\label{eq15}
g_{\mu\nu} =\begin{pmatrix}
\gamma_{ij}+ \phi A_{i}A_{j} & \quad  \phi A_{j}\\
\phi A_{i} & \phi
\end{pmatrix},
\end{equation}
where $\gamma_{ij}$ is a $3$-dimensional Riemannian metric. The inverse metric is then given by 
\begin{equation}\label{eq16}
g^{\mu\nu}=\begin{pmatrix}
\gamma^{ij} &  -A^{j}\\
-A^{i} & \quad \frac{1}{\phi}+A_{k}A^{k}
\end{pmatrix},
\end{equation}
where $\gamma^{ij}$ are the components of the inverse of $\gamma_{ij}$ and $A^{i}=\gamma^{ij}A_{j}$. We, now perform the canonical transformation explicitly given in \cite{r3}. Where $p^{ij}$, $p^{i}$ and $p$ are the conjugate momenta of $\gamma_{ij}$, $A_{i}$ and $\phi$ respectively. Then, the Hamiltonian is 
\begin{equation}\label{eq17}
\begin{split}
\mathcal{H}= \frac{N}{\sqrt{\gamma\phi}}\bigg[\phi^{2}p^{2}+p^{ij}p_{ij}+\frac{p^{i}p_{i}}{2\phi}-\gamma\phi\beta\mathbf{R}-\gamma\phi\alpha a_{i}a^{i}-\mathcal{V}\left(\gamma_{ij},A_{i},\phi,N\right)\bigg] 
-\Lambda \partial_{i}p^{i} &\\-\Lambda_{j}\bigg(2\nabla_{i}p^{ij}
+p^{i}\gamma^{jk}F_{ik}  -p\gamma^{ij}\partial_{i}\phi\bigg)-\sigma P_{N}-\mu P,
\end{split}
\end{equation}
where the four dimensional Ricci scalar $\mathds{R}$ has the usual decomposition
\begin{equation}\label{eq18}
\mathbf{R} =R-\frac{\phi}{4}F_{ij}F^{ij}-\frac{2}{\sqrt{\phi}}\nabla_{i}\nabla^{i}\sqrt{\phi}.
\end{equation}
Notice that $\mathcal{V}(g_{\mu\nu},N)$ was defined as a scalar under spacelike diffeomorphisms and time reparametrizations, 
\begin{equation}\label{eq19}
\delta \mathcal{V}=\xi^{\rho}\partial_{\rho}\mathcal{V}+f\dot{\mathcal{V}}.   
\end{equation}
Under the cylindrical projection $\partial_{4}=0$ this condition reduces to 
\begin{equation}\label{eq20}
\delta \mathcal{V}=\xi^{i}\partial_{i}\mathcal{V}+f\dot{\mathcal{V}}. 
\end{equation}
it transforms as a scalar under the symmetries of the $3+1$ foliation. We have now a $3+1$ theory invariant under the diffeomorphisms on the 3-dimensional spacelike leaves and under time reparametrization. Since we are interested in obtaining a theory where only the gravitational and electromagnetic degrees of freedom propagate, we consider at the level of the canonical action the dilaton field to be in its ground state $\phi=1$, $p=0$. The reduced theory is still invariant under diffeomorphisms on the spacelike leaves and time reparametrization since $\phi$ was a scalar field and $p$ a scalar density under diffeomorphisms on the spacelike leaves and  both scalars under reparametrizations of time. We end up with the $3+1$ anisotropic Hamiltonian 
\begin{equation}\label{eq21}
\begin{split}
\mathcal{H}= \frac{N}{\sqrt{\gamma}}\bigg[p^{ij}p_{ij}+\frac{p^{i}p_{i}}{2}-\gamma\beta R+\frac{\gamma\beta}{4}F_{ij}F^{ij}-\gamma\alpha a_{i}a^{i}
-\mathcal{V}\left(\gamma_{ij},A_{i},N\right)\bigg]
-\Lambda \tilde{H} -\Lambda_{j}H^{j}&\\-\sigma P_{N}-\mu P,
\end{split}
\end{equation}
where we have introduced the Lagrange multipliers
\begin{equation}\label{eq22}
\Lambda\equiv N_{4} \quad \mbox{and} \quad \Lambda_{i}\equiv N_{i}-A_{i}N_{4}.   
\end{equation}
The primary constraints are
\begin{eqnarray}\label{eq23}
P\equiv \gamma_{ij}p^{ij}=0, \\ \label{eq24}
P_{N}=0,
\end{eqnarray}
and the reduction of the momentum constraint (\ref{eq14}), which now decomposes into 
\begin{eqnarray}\label{eq25}
\tilde{H}\equiv\partial_{i}p^{i}&=&0 \\ \label{eq26}
H^{j}\equiv 2\nabla_{i}p^{ij}+p^{i}\gamma^{jk}F_{ik}&=&0.
\end{eqnarray}
It is worth mentioning that (\ref{eq25}) is exactly the same first class constraint associated to the $U(1)$ gauge symmetry in the relativistic electromagnetic theory, while (\ref{eq26}) is the generator of spacelike diffeomorphisms on the $3+1$ foliation. In order to determine the field equations of the theory at low energies,  we dismiss the higher order spacelike derivative terms on the potential $\mathcal{V}$. The conservation of the primary constraints (\ref{eq23})-(\ref{eq24}) yields 
\begin{equation}\label{eq27}
\begin{split}
H_{P}\equiv\frac{3}{2}\frac{1}{\sqrt{\gamma}}p^{lm}p_{lm}+\frac{1}{4}\frac{1}{\sqrt{\gamma}}p^{k}p_{k}+\frac{1}{2}\sqrt{\gamma}\beta R+\frac{1}{8}\sqrt{\gamma}\beta F^{lm}F_{lm}+\sqrt{\gamma}\left(\frac{\alpha}{2}-2\beta\right)a^{k}a_{k}  &\\-2\beta\sqrt{\gamma}\nabla^{l}a_{l}=0,
\end{split}
\end{equation}
\begin{equation}\label{eq28}
\begin{split}
H_{N}\equiv\frac{1}{\sqrt{\gamma}}\bigg[p^{ij}p_{ij}+\frac{p^{i}p_{i}}{2}-\beta\gamma R +\frac{\beta}{4} \gamma F_{ij}F^{ij} \bigg]+\alpha\sqrt{\gamma}a_{i}a^{i}
+2\alpha\sqrt{\gamma}\nabla_{i}a^{i}=0.
\end{split}
\end{equation}
At this stage the Dirac algorithm to obtain the constraints of the theory ends up. In fact, conservation of (\ref{eq27}) and (\ref{eq28}) only determine Lagrange multipliers. The analysis of the constraints determines that (\ref{eq23}), (\ref{eq24}), (\ref{eq27}) and (\ref{eq29}) are second class constraints while (\ref{eq25}) and (\ref{eq26}) are first class ones. 

The transformation laws of the $3+1$ formulation are the following 
\begin{eqnarray} \label{eq29}
\delta \gamma_{ij}&=& \partial_{i}\xi^{k}\gamma_{jk}+\partial_{j}\xi^{k}\gamma_{ik}+\xi^{k}\partial_{k}\gamma_{ij}+f\dot{\gamma}_{ij}, \\ \label{eq30}
\delta \Lambda_{i}&=&\partial_{i}\xi^{k}\Lambda_{k}+\xi^{k}\partial_{k}\Lambda_{i}+\dot{\xi}^{k}\gamma_{ki}+f\dot{\Lambda}_{i}+\dot{f}\Lambda_{i}, \\ \label{eq31}
\delta N&=& \xi^{k}\partial_{k}N+f\dot{N}+\dot{f}N,\\ \label{eq32}
\delta A_{i}&=&\partial_{i}\xi^{k}A_{k}+\xi^{k}\partial_{k}A_{i}+f\dot{A}_{i}+\partial_{i}\xi^{4}, \\ \label{eq33}
\delta\Lambda&=&\xi^{i}\partial_{i}\Lambda+\dot{\xi}^{i}A_{i}+\dot{\xi}^{4}+f\dot{\Lambda}+\dot{f}\Lambda.
\end{eqnarray}
We observe that (\ref{eq29})-(\ref{eq31}) are the same infinitesimal transformation laws as for the $3+1$ dimensional Ho\v{r}ava-Lifshitz gravity theory. While (\ref{eq32}) is the transformation law for a gauge vector.
Next, the dynamic of the theory is described by the following field equations 
\begin{eqnarray}\label{eq34}
\dot{\gamma}_{ij}&=&\frac{2N}{\sqrt{\gamma}}p_{ij}+\nabla_{i}\Lambda_{j}+\nabla_{j}\Lambda_{i}+\mu \gamma_{ij}, \\ \label{eq35}
\dot{A}_{i}&=&\frac{N}{\sqrt{\gamma}}p_{i}+\partial_{i}\Lambda-\Lambda_{j}\gamma^{jk}F_{ik},
\end{eqnarray}
\begin{equation}\label{eq36}
\begin{split}
\dot{p}^{ij}=\frac{N}{2}\frac{\gamma^{ij}}{\sqrt{\gamma}}\left[p^{lk}p_{lk}+\frac{p^{l}p_{l}}{2}\right]  
-\frac{N}{\sqrt{\gamma}}\left[2p^{ik}p^{j}_{k}+\frac{p^{i}p^{j}}{2}\right]
+N\sqrt{\gamma}\beta\left[\frac{R}{2}\gamma^{ij}
- R^{ij}\right]&\\+\beta\sqrt{\gamma}\bigg[\nabla^{(i}\nabla^{j)}N-\gamma^{ij}\nabla_{k}\nabla^{k}N\bigg]
+\frac{\beta}{2}N\sqrt{\gamma}\Bigg[F^{in}F^{\ j}_{n}
-\frac{\gamma^{ij}}{4}F_{mn}F^{mn}\Bigg]
&\\+\alpha N\sqrt{\gamma} \bigg[\frac{\gamma^{ij}}{2}a_{k}a^{k}
-a^{i}a^{j}\bigg]
-\nabla_{k}\bigg[2p^{k(i}\Lambda^{j)}-p^{ij}\Lambda^{k}\bigg]-\Lambda^{(i}\gamma^{j)m}p^{l}F_{lm}+\mu p^{ij},
\end{split}    
\end{equation}
\begin{equation}\label{eq37}
\dot{p}^{i}=\beta\partial_{j}\left(N\sqrt{\gamma} F^{ji}\right)+\partial_{k}\left(\Lambda^{k}p^{i}-\Lambda^{i}p^{k}\right).
\end{equation}
It should be noted that two coupling constants are involved in the dynamics in comparison to Einstein-Maxwell theory, $\beta$ and $\alpha$. Nevertheless, the equations describing the dynamics of the gauge vector that is (\ref{eq25}), (\ref{eq35}) and (\ref{eq37}) only involve the coupling $\beta$. If $\beta=1$
this three equations are exactly the same ones as in the Einstein-Maxwell relativistic theory. 
In section \ref{sec6} we will compare the anisotropic electromagnetic-gravitational theory deduced in this section with the Einstein-Maxwell relativistic one. In order to improve the understanding of the dynamics of the anisotropic theory we perform in the next section a perturbative analysis of the field equations around a background consisting on an Euclidean $3$-dimensional  metric with no electromagnetic interaction.

\section{Perturbative Analysis: Propagation of Degrees of Freedom}\label{sec4}
In order to unravel the propagating physical degrees of freedom, We perform in this section a perturbative analysis. To do so we introduce perturbations around an Euclidean with not electromagnetic interaction background in the following way
\begin{equation}\label{eq38}
\gamma_{ij}=\delta_{ij}+\epsilon h_{ij}, \quad p^{ij}=\epsilon\Omega_{ij}, 
\end{equation}
\begin{equation}
 \Lambda_{i}=\epsilon n_{i}, \quad \Lambda=\epsilon n_{4}, \quad N=1+\epsilon n,    
\end{equation}
and for the vector field $A_{i}$ we have
\begin{equation}\label{eq39}
A_{i}=\epsilon\xi_{i}, \quad p^{i}=\epsilon\zeta_{i}.
\end{equation}
Therefore, at linear order in $\epsilon$ the field equations (\ref{eq34})-(\ref{eq37}) become
\begin{equation}\label{eq40}
\dot{h}_{ij}=2\Omega_{ij}+2\partial_{(i}n_{j)}+\mu\delta_{ij}, 
\end{equation}
\begin{equation}\label{eq41}
\begin{split}
\dot{\Omega}_{ij}=-\frac{\beta}{2}\left(\delta_{ij}-\frac{\partial_{i}\partial_{j}}{\Delta}\right)\Delta h+\frac{\beta}{2}\Delta h_{ij}-\beta\left(\delta_{ij}-\frac{\partial_{i}\partial_{j}}{\Delta}\right)\Delta n &\\
+\frac{\beta}{2}\left(-\partial_{j}\partial_{k}h_{ik}-\partial_{i}\partial_{k}h_{jk}+\delta_{ij}\partial_{l}\partial_{k}h_{lk}\right),
\end{split}
\end{equation}
\begin{eqnarray}\label{eq42}
\dot{\xi}_{i}&=&\zeta_{i}+\partial_{i}n_{4}, \\ \label{eq43}
\dot{\zeta}_{i}&=&\beta\partial_{j}\left(\partial_{j}\xi_{i}-\partial_{i}\xi_{j}\right).
\end{eqnarray}
Furthermore, from the constraints (\ref{eq26}), (\ref{eq27}), (\ref{eq28}) and (\ref{eq29}) we have
\begin{eqnarray}\label{photon}
\partial_{i}\zeta_{i}&=&0 \\ \label{eq44}
\partial_{i}\Omega_{ij}&=&0\\ \label{eq45}
\left(\alpha-\beta\right)\Delta n&=&0 \\ \label{new}
\beta\left(\partial_{i}\partial_{j}h_{ij}-\Delta h\right)&=&0.
\end{eqnarray}
The content of equations
(\ref{eq40})-(\ref{new}) can be seen more easily if one makes the ADM orthogonal transverse/longitudinal decomposition on $h_{ij}$, $\Omega^{ij}$, $\xi_{i}$ and $\zeta^{i}$, obtaining, if $\alpha-\beta$ and $\beta$ are different from zero, $n=0$, $h^{T}=0$ and
\begin{equation}\label{eq46}
\dot{\xi}^{T}_{i}=\zeta^{T}_{i}.    
\end{equation}
\begin{equation}\label{eq47}
\dot{\zeta}^{T}_{i}=\beta\Delta\xi^{T}_{i},    
\end{equation}
for the gauge vector field equations. Then, combining (\ref{eq46}) and (\ref{eq47}) we get the following wave equation for the gauge vector
\begin{equation}\label{eq48}
\ddot{\xi}^{T}_{i}-\beta\Delta\xi^{T}_{i}=0. 
\end{equation}
From equations (\ref{eq40}) and (\ref{eq41}) we obtain the following wave equation for the graviton
\begin{equation}\label{eq49}
\ddot{h}^{TT}_{ij}-\beta\Delta h^{TT}_{ij}=0.   
\end{equation}
We remark that the only propagating degrees of freedom correspond to the transverse traceless gravity modes and the transverse electromagnetic modes, exactly as in the Einstein-Maxwell theory. By working at the kinetic conformal point we got rid of the scalar Ho\v{r}ava-Lifshitz mode and by working on the ground state of the dilaton field we got rid of the Kaluza-Klein scalar mode, we are then left with only the propagating modes of the two fundamental interactions, \i.e gravity and electromagnetic but now in an anisotropic, non relativistic scenario. In particular, the gravity and electromagnetic modes propagate at the same speed $\sqrt{\beta}$. There is no propagation of scalar fields, hence there is no decoupling problem and the theory will flow from the UV point to the IR one with the same physical degrees of freedom. Moreover, the longitudinal modes of $h_{ij}$ are gauge modes which can be fixed to zero. The same occurs with the longitudinal modes of the gauge vector. The transverse mode of the metric was eliminated from one of the second class constraints. The  transverse  $\Omega^{T}$ mode together with the longitudinal components of $\Omega_{ij}$ were eliminated  from the constraints. Only the transverse traceless gravitational modes and the transverse electromagnetic modes propagate, the other ones are zero at first order in perturbations or are gauge modes which can be taken to be zero. 

\section{The anisotropic electromagnetic theory}\label{sec5}
In the previous sections we obtained the anisotropic electro-
magnetic-gravitational theory \emph{a la} Ho\v{r}ava-Lifshitz (EGHL theory). In order to eliminate the scalar field of the Ho\v{r}ava-Lifshitz gravity we started with the Ho\v{r}ava-Lifshitz in $4+1$ dimensions at the critical value of the dimensionless coupling constant $\lambda=1/4$, the kinetic conformal theory. This value of $\lambda$ is protected  from quantum corrections by the second class constraint $P=0$. We then performed a Kaluza-Klein reduction to $3+1$ dimensions and obtained a theory describing the evolution of the 3-dimensional metric $\gamma_{ij}$ of the spacelike leaves of the foliation, a gauge vector $A_{i}$, the dilaton scalar field, the lapse and shift fields describing the embedding of the leaves in the $4+1$ manifold. We then considered the dilaton scalar field on its ground state, $\phi=1$ and its conjugate momentum $p=0$. The resulting EGHL theory is consistent and propagates only the electromagnetic and gravitational interactions, no scalar fields are present. If we impose $A_{i}=0$, $p^{i}=0$ on the field equations of the EGHL theory we obtain the field equations of the Ho\v{r}ava-Lifshitz gravity theory at the kinetic conformal point \cite{r4,r5}.
An important point to consider is the comparison of the anisotropic electromagnetic equations with the relativistic Maxwell equations, since the latter are very well established. The equations for the anisotropic electromagnetic interaction are (\ref{eq25}), (\ref{eq35}) and (\ref{eq37}). Given a metric $\gamma_{ij}$, the lapse $N$ and shift $\Lambda_{i}$ these equations depend only the coupling constant $\beta$. The dependence on the coupling $\alpha$ is through $\gamma_{ij}$, $N$ and $\Lambda_{i}$. In order to compare the anisotropic electromagnetic equations with Maxwell equations we introduce the $3+1$ metric associated to the foliation. We notice from the transformation law (\ref{eq30}) that $\Lambda_{i}=N_{i}-N_{4}A_{i}$ is the correct shift, which together with the lapse $N$ describe the embedding of the leaves in the $3+1$ foliation. Notice that $N_{i}$ does not have the transformation law of the shift in the $3+1$ foliation. We introduce then the $3+1$ metric 
\begin{equation}\label{eq54}
g_{\mu\nu} =\begin{pmatrix}
-N^{2}+\Lambda_{k}\Lambda^{k} &  \Lambda_{i}\\
\Lambda_{j} & \quad \gamma_{ij}
\end{pmatrix},
\end{equation}
with inverse
\begin{equation}\label{eq55}
g^{\mu\nu}=\begin{pmatrix}
-\frac{1}{N^{2}} &  \frac{\Lambda^{i}}{N^{2}}\\
\frac{\Lambda^{j}}{N^{2}} & \quad \gamma^{ij}-\frac{\Lambda^{i}\Lambda^{j}}{N^{2}}
\end{pmatrix},
\end{equation}
where $\Lambda^{j}=\gamma^{ij}\Lambda_{i}$, and $\gamma^{ij}$ is the inverse of $\gamma_{ij}$. We then obtain 
\begin{equation}\label{eq56}
^{4}F^{0i}\equiv g^{0\mu}g^{i\nu}F_{\mu\nu}=-\frac{1}{N^{2}}\gamma^{ik}F_{0k}+\frac{\gamma^{ik}}{N^{2}}\Lambda^{m}F_{mk},    
\end{equation}
where $F_{\mu\nu}=\partial_{\mu}A_{\nu}-\partial_{\nu}A_{\mu}$, $\mu=0,1,2,3$ and we have introduced $\Lambda\equiv A_{0}$ the Lagrange multiplier associated to the first class constraint (\ref{eq25}), the generator of gauge transformations. So, from (\ref{eq35}) we get 
\begin{equation}\label{eq57}
\gamma^{ik}F_{0k}=\frac{N}{\sqrt{\gamma}}p^{i}-\Lambda^{m}F_{mk}\gamma^{ik}.  
\end{equation}
By replacing the left hand member of this equation into (\ref{eq56}) we have 
\begin{equation}\label{eq58}
p^{i}=-N\sqrt{\gamma}\ ^{4}F^{0i},    
\end{equation}
where $N\sqrt{\gamma}=\sqrt{g}$. We use now
\begin{equation}\label{eq59}
^{4}F^{ji}\equiv g^{j\mu}g^{i \nu}F_{\mu\nu}=g^{j0}g^{ik}F_{ok}+g^{jk}g^{i0}F_{k0}+g^{jm}g^{in}F_{mn},
\end{equation}
and after some calculations, where we used equation (\ref{eq57}), we get 
\begin{equation}\label{eq60}
^{4}F^{ij}=\frac{1}{N\sqrt{\gamma}}\left[\Lambda^{i}p^{j}-\Lambda^{j}p^{i}\right]+\gamma^{im}\gamma^{jn}F_{mn}.    
\end{equation}
From (\ref{eq37}) and the previous equation we arrive to
\begin{equation}\label{eq61}
\dot{p}^{i}=\beta\partial_{j}\left[N\sqrt{\gamma}\ ^{4}F^{ji}\right]+\left(1-\beta\right)\partial_{j}\left[\Lambda^{i}p^{j}-\Lambda^{j}p^{i}\right].    
\end{equation}
Also, from (\ref{eq25}) and (\ref{eq58}) we obtain
\begin{equation}\label{eq62}
\partial_{i}\left(\sqrt{g}\ ^{4}F^{i0}\right)=0,
\end{equation}
which is equivalent to  
\begin{equation}\label{new2}
\nabla_{\mu} \ ^{4}F^{\mu0}=0.    
\end{equation}
Finally,, from (\ref{eq58}) and (\ref{eq61}) we have
\begin{equation}\label{eq63}
\nabla_{\mu}\ ^{4}F^{\mu i}+\frac{\left(1-\beta\right)}{\sqrt{g}}\partial_{j}\left[\Lambda^{j}p^{i}-\Lambda^{i}p^{j}+\sqrt{g}\ ^{4}F^{ij}\right]=0.    
\end{equation}
These are the field equations for the anisotropic electromagnetic theory. The deviation from Maxwell equations is proportional to $\left(1-\beta\right)$. If $\beta=1$, (\ref{eq63}) reduces to the Maxwell equations. Hence, if the renormalization flow reaches $\beta=1$ at the infrared point, we would recover the Maxwell electromagnetic theory. Besides for any value of $\beta$, if we impose the gauge condition $\Lambda^{j}=0$, an admissible gauge condition since $\Lambda_{j}$ is the Lagrange multiplier associated to the first class constraint $H^{j}=0$, the field equations have the same form as Maxwell equations but with a propagating speed $\sqrt{\beta}$.
\section{Equivalence to the Einstein-Maxwell theory}\label{sec6}
We consider now the theory with the coupling constants $\beta=1$ and $\alpha=0$. We will prove that this theory is the same as the Einstein-Maxwell theory in a particular gauge. We are considering only up to quadratic derivative terms in the potential, the low energy limit of the theory. Since the theory is restricted by the constraint $P=0$, one may add to the Lagrangian any term proportional to $P$. We can add in (\ref{eq21}) a term $-\frac{1}{2}P^{2}$. The Hamiltonian without the $\mu P$ term is exactly the Einstein-Maxwell Hamiltonian,
\begin{equation}\label{eq64}
\mathcal{H}=\mathcal{H}_{E-M}-\mu P,    
\end{equation}
where the Einstein-Maxwell Hamiltonian $\mathcal{H}_{E-M}$ is
\begin{equation}\label{eq65}
\begin{split}
\mathcal{H}_{E-M}=\frac{N}{\sqrt{\gamma}}\bigg[p^{ij}p_{ij}-\frac{P^{2}}{2}+\frac{p^{i}p_{i}}{2}-\gamma R+\frac{\gamma}{4}F_{ij}F^{ij}\bigg] 
-\Lambda \tilde{H}  -\Lambda_{j}H^{j}. 
\end{split}
\end{equation}
The constraints (\ref{eq25}) and (\ref{eq26}) are the same as in Einstein-Maxwell theory. Furthermore (\ref{eq28}) with the inclusion of the $-\frac{1}{2}P^{2}$ term and $\alpha=0$, becomes the Hamiltonian constraint of the Einstein-Maxwell theory, a first class constraint in that theory but not in the present analysis because of the constraint $P=0$.
\begin{equation}\label{eq66}
\begin{split}
H_{N}\equiv\frac{1}{\sqrt{\gamma}}\bigg[p^{ij}p_{ij}-\frac{P^{2}}{2}+\frac{p^{i}p_{i}}{2}-\frac{1}{2}P^{2}-\gamma R +\frac{\gamma}{4} F_{ij}F^{ij} \bigg]=0.
\end{split}    
\end{equation}
The conservation of the Hamiltonian constraint determines the value of $\mu$. It is straightforward to show that $\mu=0$. In fact $H_{N}$ weakly commutes under the Poisson bracket with $\mathcal{H}_{E-M}$. Equations (\ref{eq35}) and (\ref{eq36}) are then the same as in Einstein-Maxwell theory
\begin{eqnarray}\label{eq67}
\dot{\gamma}_{ij}&=&\frac{2N}{\sqrt{\gamma}}\left[p_{ij}-\frac{\gamma_{ij}}{2}P\right]+\nabla_{i}\Lambda_{j}+\nabla_{j}\Lambda_{i}, \\ \label{eq68}
\dot{A}_{i}&=&\frac{N}{\sqrt{\gamma}}p_{i}+\partial_{i}\Lambda-\Lambda_{j}\gamma^{jk}F_{ik}.
\end{eqnarray}
Also, equations (\ref{eq37}) and (\ref{eq38}) with the inclusion of the terms proportional to $P$ arising from the Hamiltonian and using $\mu=0$ are the same field equations as in Einstein-Maxwell theory,
\begin{equation}\label{eq69}
\begin{split}
\dot{p}^{ij}=\frac{N}{2}\frac{\gamma^{ij}}{\sqrt{\gamma}}\left[p^{lk}p_{lk}-\frac{P^{2}}{2}+\frac{p^{l}p_{l}}{2}\right]  
-\frac{N}{\sqrt{\gamma}}\bigg[2p^{ik}p^{j}_{k}-p^{ij}P+\frac{p^{i}p^{j}}{2}\bigg]
+N\sqrt{\gamma}\left[\frac{R}{2}\gamma^{ij}
- R^{ij}\right]&\\
+\sqrt{\gamma}\bigg[\nabla^{(i}\nabla^{j)}N-\gamma^{ij}\nabla_{k}\nabla^{k}N\bigg]
+\frac{N}{2}\sqrt{\gamma}\Bigg[F^{in}F^{\ j}_{n}
-\frac{\gamma^{ij}}{4}F_{mn}F^{mn}\Bigg]
-\nabla_{k}\bigg[2p^{k(i}\Lambda^{j)}-p^{ij}\Lambda^{k}\bigg]&\\
-\Lambda^{(i}\gamma^{j)m}p^{l}F_{lm},
\end{split}    
\end{equation}
\begin{equation}\label{eq70}
\dot{p}^{i}=\partial_{j}\left(N\sqrt{\gamma} F^{ji}\right)+\partial_{k}\left(\Lambda^{k}p^{i}-\Lambda^{i}p^{k}\right).
\end{equation}
The whole set of field equations of the anisotropic electromag-
netic-gravitational theory agree exactly with the ones of Eins-
tein-Maxwell theory, once one imposes the partial gauge fixing condition in the Einstein-Maxwell theory 
\begin{equation}\label{eq71}
P=0,    
\end{equation}
a gravitational gauge used in the ADM formulation of General Relativity \cite{r26}. The conservation of it renders equation (\ref{eq27}).

We may now consider the path integral formulation of both theories, where only quadratic terms in the potential of the EGHL theory are included and $\alpha=0$ and $\beta=1$. The measure of the path integral formulation of the Einstein-Maxwell theory in above mentioned gauge is
\begin{equation}
\begin{split}
d\mu\left(\gamma_{ij},p^{lk},A_{m},p^{n}\right)=\prod \delta\left(\tilde{H}\right)\delta\left(\chi\right)det|\{\tilde{H},\chi\}_{PB}|\delta\left(H^{j}\right) \delta\left(\chi_{i}\right)det|\{H^{j},\chi_{i}\}_{PB}|&\\ \times\delta\left(H_{N}\right)\delta\left(P\right)   det|\{H_{N},P\}_{PB}|\prod d\gamma_{ij}dp^{lk}dA_{m}dp^{n}.
\end{split}
\end{equation}
The measure of the path integral formulation of the EGHL theory is
\begin{equation}
\begin{split}
d\tilde{\mu}\left(\gamma_{ij},p^{lk},A_{m},p^{n},N,P_{N}\right)=\prod \delta\left(\tilde{H}\right)\delta\left(\chi\right)det|\{\tilde{H},\chi\}_{PB}| \delta\left(H^{j}\right)\delta\left(\chi_{i}\right)det|\{H^{j},\chi_{i}\}_{PB}| & \\ \times \delta\left(\theta_{1}\right)\delta\left(\theta_{2}\right)\delta\left(\theta_{3}\right)\delta\left(\theta_{4}\right)\left(det|\{\theta_{i},\theta_{j}\}_{PB}|\right)^{1/2}\prod d\gamma_{ij}dp^{lk}dA_{m}dp^{n}dNdP_{N},
\end{split}    
\end{equation}
where $\theta_{1}\equiv H_{N}$, $\theta_{2}\equiv P$, $\theta_{3}\equiv H_{P}$ and $\theta_{4}\equiv P_{N}$. 

In the Einstein-Maxwell theory, $\tilde{H}=0$,  $H^{j}=0$ and $H_{N}=0$ are first class constraints while in the EGHL theory, $\tilde{H}=0$ and  $H^{j}=0$ are first class constraints and $\theta_{i}=0$, $i=1,...,4$ are second class constraints. $\chi_{i}=0$,  $i=1,...,4$ and $\chi=0$ are the gauge fixing conditions associated to the first class constraints $H^{j}=0$ and $\tilde{H}$. The equivalence of the measure occurs because 
\begin{equation}
\left(det|\{\theta_{i},\theta_{j}\}_{PB}|\right)^{1/2}=det|\{\theta_{1},\theta_{2}\}_{PB}|det|\{\theta_{3},\theta_{4}\}_{PB}|.
\end{equation}
Moreover
\begin{equation}
\delta\left(\theta_{3}\right)\delta\left(\theta_{4}\right)det|\{\theta_{3},\theta_{4}\}_{PB}|=\delta\left(N-\hat{N}\right)\delta\left(P_{N}\right),    
\end{equation}
where $\hat{N}$ is the solution of the elliptic equation (\ref{eq27}). Hence after integration of $N$ and $P_{N}$ we get the equality of both path integral formulations, taking into account that the Hamiltonian is the same in both theories.

\section{Concluding remarks}\label{sec7}
We introduced an anisotropic theory in the sense of Ho\v{r}ava-Lifshitz theory describing the electromagnetic-gravitational interaction. The propagating physical degrees of freedom are only the transverse-traceless gravity modes together with the transverse electromagnetic modes as in the Einstein-Max-
well relativistic theory, there are no propagating scalar degrees of freedom in the formulation. In the infrared limit the theory depends on two coupling constant $\alpha$ and $\beta$. The dimensionless coupling $\lambda$ of the kinetic term in the Ho\v{r}ava-Lifshitz gravity theory, which for $\lambda\neq 1$ breaks  the relativistic symmetry of the kinetic term, takes its critical value $\lambda=1/4$. The kinetic term becomes conformal invariant at that value of $\lambda$, however the potential terms in the action break that symmetry. The Ho\v{r}ava-Lifshitz gravity theory at the critical point $\lambda=1/3$ was studied in \cite{r4,r5}.

The anisotropic electromagnetic-gravitational theory shift the critical value to $\lambda=1/4$. That value of $\lambda$ is protected of quantum corrections, by a second class constraint, the trace of the conjugate momentum of the 3-dimensional metric of the spacelike leaves of the foliation, $P=0$. This constraint, which ends up being a second class one has not been added by hand, but arises directly as a primary constraint of the theory following the Dirac approach to determine the whole set of constraints. The first class constraints generates the diffeomorphisms on the spacelike leaves and the local gauge transformations of the anisotropic electromagnetic interaction associated to a $U(1)$ principle bundle with a connection one-form, the gauge vector. Besides the second class constraint already mentioned $P=0$, the Hamiltonian constraint is also a second class constraint, in distinction to what occurs in the Einstein-Maxwell theory. The conservation of $P=0$ yields an additional second class constraint, which can be interpreted as an elliptic partial differential equation for the lapse $N(x,t)$. the Hamiltonian constraint is an elliptic partial differential equation for $g^{T}$, in the ADM notation. $P^{T}$ is eliminated from $P=0$ while the longitudinal components of the metric, and its conjugate momentum together with the longitudinal component of the gauge vector and its conjugate momentum are eliminated from the first class constraint and its gauge fixing conditions.
For any value of $\beta$ and $\alpha$ the anisotropic electromagnetic-gravitational theory propagates the same physical degrees of freedom as the Einstein-Maxwell theory. To show it, we performed a perturbative analysis on a Euclidean background. The propagating physical degrees of freedom are the transverse traceless components of the 3-dimensional spacelike metric and the transverse modes of the gauge vector. This is so because we are at the critical value of the coupling $\lambda$, the kinetic conformal point. The intrinsic scalar field of the Ho\v{r}ava-Lifshitz formulation is not present in that case. Besides we consider the dilaton scalar field arising from the Kaluza-Klein approach to be in its ground state. All excitations propagate with the same speed $\sqrt{\beta}$.

We showed that the complete algebraic structure of the constraints is consistent and in agreement with the transformation law of the geometrical objects of the theory. We derived the Hamiltonian and field equations of the theory and compared them to the field equations of the Einstein-Maxwell theory. We emphasize that the theory is a non-projectable one. The projectable version of the theory  although an interesting one propagates a different number of physical degrees of freedom compared to the non-projectable one.

Furthermore, we showed that for the coupling constants with value $\beta=1$ the field equations for the gauge vector are exactly the Maxwell equations on a gravitational background characterized by 3-dimensional metric of the spacelike leaves, the lapse and the shift. The interesting point is that the shift acquires a contribution from the gauge vector which renders the correct transformation law for it. The dependence on the coupling constant $\alpha$ is only through the background fields. 

We also obtained the general field equations for the gauge vector for any value of $\beta$, it is a deviation of Maxwell equations by terms proportional to $\left(1-\beta\right)$. In the gauge $\Lambda_{i}=0$, the field equations have the same form as Maxwell equations but with a speed of propagation $\sqrt{\beta}$.

If the coupling constants take the values $\beta=1$ and $\alpha=0$ we proved that the anisotropic electromagnetic-gravitational theory is exactly the Einstein-Maxwell theory on a particular gauge used in the ADM formulation of General Relativity \cite{r26}. This result is an extension of the analogous one for the Ho\v{r}ava-Lifshitz gravity at the kinetic conformal point \cite{r5}. Moreover, the path integral formulation of this anisotropic theory when only up to second spacelike derivatives are included in the potential, is exactly the same functional defining the path integral formulation of the Einstein-Maxwell theory (provided there are no gauge anomalies). The path integral of the quantum formulation of the anisotropic electro-
magnetic-gravitational theory also includes all higher derivative terms in the potential. It introduces in a natural way all higher derivative terms, up to $z=4$ terms, which modify the Eisntein-Maxwell theory in the UV regime.  

The virtue of the anisotropic electromagnetic-gravitatio-
nal theory we consider is that we started with a Ho\v{r}ava-Lifshitz gravity theory in $4+1$ dimensions, which is power counting renormalizable theory. We ended up with a $3+1$ theory with all higher order spacelike derivative terms, up to $z=4$, and with the geometrical structure and symmetries of the Ho\v{r}ava-Lifshitz approach. We then expect the theory to be power counting renormalizable. This problem will be considered elsewhere.

Our construction differs with what is usually done to describe the electromagnetic interaction coupled to the Ho\v{r}ava-Lifshitz gravity theory, where the relativistic electromagnetic action is coupled to the Ho\v{r}ava-Lifshitz gravity action without the inclusion of higher derivative terms for the electromagnetic sector. Consequently, in the latter approach there is no guarantee of power counting renormalization.

In our approach we started from Ho\v{r}ava-Lifshitz gravity formulation in $4+1$ dimensions and performed a Kaluza-Klein dimensional reduction to $3+1$ dimensions. There are at this point two interpretations. One of them is to take the $3+1$ formulation, including the higher order spacelike derivative terms, to be the complete theory. In that case the higher dimensional construction is only a geometrical construction which provides in a natural way the anisotropic geometry in $3+1$ dimensions. The other interpretation is that there is a more fundamental theory at higher dimensions, and the dilaton scalar field should then be incorporated in the UV regime. To consider the scalar field at its ground state in the IR regime has then a physical meaning and is not just a geometrical mechanism. In this case the theory should determine at what energy scale the dilaton field decouples, an old problem in string theory. In this paper we take the first point of view, hence there are no propagating scalar degrees of freedom in the $3+1$ theory for any value of the couplings. Consequently, in the renormalization flow from UV point to the IR one there does not exist any decoupling problem. The main point is then to prove renormalizability of the theory. We expect to discuss aspects of it elsewhere.

\section*{ACKNOWLEDGEMENTS}
A. Restuccia and F. Tello-Ortiz are partially supported by grant Fondecyt No. $1161192$, Chile. F. Tello-Ortiz thanks the financial support by the CONICYT PFCHA/DOCTORADO-NACIONAL/2019-21190856 and project ANT-1856 at the Universidad de Antofagasta, Chile. F. Tello-Ortiz also thanks the fruitful discussions with Dr. Tom\'as Ort\'in during his stay at the Institute for Theoretical Physics (IFT) Madrid, Spain.


\end{document}